\documentclass{iau} 
\usepackage{graphicx}

\title[Kinematics of Multiple Stellar Populations] 
{Kinematics of Multiple Stellar Populations in Globular Clusters with Gaia}

\author[G.\,Cordoni et al.]   
{G.\,Cordoni$^1$, A.~P.\,Milone$^1$, A.\,Mastrobuono-Battisti$^2$, A.~F.\,Marino$^{1,3}$, E.~P.\,Lagioia$^1$ \and M.\,Tailo$^1$}

\affiliation{$^1$Dipartimento di Fisica e Astronomia ``Galileo Galilei'' -
  Univ. di Padova \\ Vicolo dell'Osservatorio 3, Padova, IT-35122  \\[\affilskip]
$^2$Max-Planck Institut f{\"u}r Astronomie, K{\"o}nigstuhl 17, D-69117 Heidelberg, Germany \\
$^3$Centro di Ateneo di Studi e Attivit\`a Spaziali ``Giuseppe Colombo'' - CISAS \\ Via Venezia 15, Padova, IT-35131 }

\pubyear{2019}
\volume{351}  
\jname{Star Clusters: From the Milky Way to the Early Universe}
\editors{A. Bragaglia, M.B. Davies, A. Sills \& E. Vesperini, eds.}
\begin{document}

\maketitle
\begin{abstract}
The internal dynamics of multiple stellar populations in Globular Clusters (GCs) provides unique constraints on the physical processes responsible for their formation. Specifically, the present-day kinematics of cluster stars, such as rotation and velocity dispersion, seems to be related to the initial configuration of the system. In recent work (Milone et al. 2018), we analyzed for the first time the kinematics of the different stellar populations in NGC\,0104 (47\,Tucanae) over a large field of view, exploiting the Gaia Data Release 2 proper motions combined with multi-band ground-based photometry. In this paper, based on the work by Cordoni et al. (2019), we extend this analysis to six GCs, namely NGC\,0288, NGC\,5904 (M\,5), NGC\,6121 (M\,4), NGC\,6752, NGC\,6838 (M\,71) and further explore NGC\,0104. 
Among the analyzed clusters only NGC\,0104 and NGC\,5904 show significant rotation on the plane of the sky.  Interestingly, multiple stellar populations in NGC\,5904 exhibit different rotation curves.

\end{abstract}

\firstsection 
\section{Introduction}

  Studies based on {\it Hubble Space Telescope} ({\it HST}) images revealed that the photometric diagrams of nearly all GCs are composed of two main groups of first-generation (1G) and second-generation stars (2G, e.g. \cite[Milone et al. 2017]{milone2017}), with different chemical compositions. Many efforts have been made to understand their origin, but, so far, none of the proposed scenarios have been able to reach a satisfactory agreement with the observations. \\
  According to many of these scenarios, 2G stars formed out of the ejecta of more massive 1G stars after the segregation of the gas in the cluster center. As a consequence, 2G stars formed in a more centrally-concentrated environment, with respect to that of 1G stars. \\
  As an alternative hypothesis, GCs host a single stellar generation and stars with different chemical composition are the product of exotic physical phenomena specific of proto-GCs. \\
  An important signature of the physical processes responsible for the formation of multiple populations is the kinematics of cluster stars. Specifically, $N$-body simulations suggest that the dynamical evolution of more centrally-concentrated 2G stars should be significantly different from that of 1G stars, and such difference could still be observable in present-day GC kinematics (\cite[Vesperini et al. 2013, H{\'e}nault-Brunet et al. 2015, Mastrobuono-Battisti \& Perets 2016]{vesperini2013, henault2015,mastrobuono2016}). \\
  In the past decade, nearly all works on the kinematics of GCs were based on radial velocities of a relatively-small sample of stars. 
  More recently, {\it HST} provided high-precision relative proper motions of a small but increasing number of clusters, namely NGC\,0104 (47\,Tucanae), NGC\,0362, NGC\,2808, NGC\,5139 and NGC\,6352 that allowed the investigation of the kinematics of multiple populations on the plane of the sky (e.g. \cite[Richer et al. 2013, Bellini et al. 2015]{richer2013, bellini2015}). 
  While these works are based on high-precision relative proper motions of thousands of stars, the small field of view of {\it HST} does not allow the study of the entire cluster. 
  To overcome this shortcoming and study the kinematics of multiple stellar populations over a large field of view, we started a project based on Gaia Data Release 2 (\cite[Gaia DR2, Gaia Collaboration et al. 2018]{gaia2018a}) accurate proper motions and multi-band wide-field ground-based photometry (\cite[Stetson et al. 2019]{stetson2019}).
  In the pilot paper of this project, we investigated for the first time the kinematics of 1G and 2G stars of NGC\,0104 over a wide field of view, up to $\sim$18 arcmin from the cluster center (corresponding to $\sim 22$\,pc, \cite[Milone et al. 2018]{milone2018}).
  In this work, we further analyse NGC\,0104 and extend the study to other six Galactic GCs, namely NGC\,0288, NGC\,5904 (M\,5), NGC\,6121 (M\,4), NGC\,6254 (M\,10), NGC\,6752 and NGC\,6838 (M\,71).

\begin{figure}[h!]
  \centering
  \includegraphics[width=6.5cm]{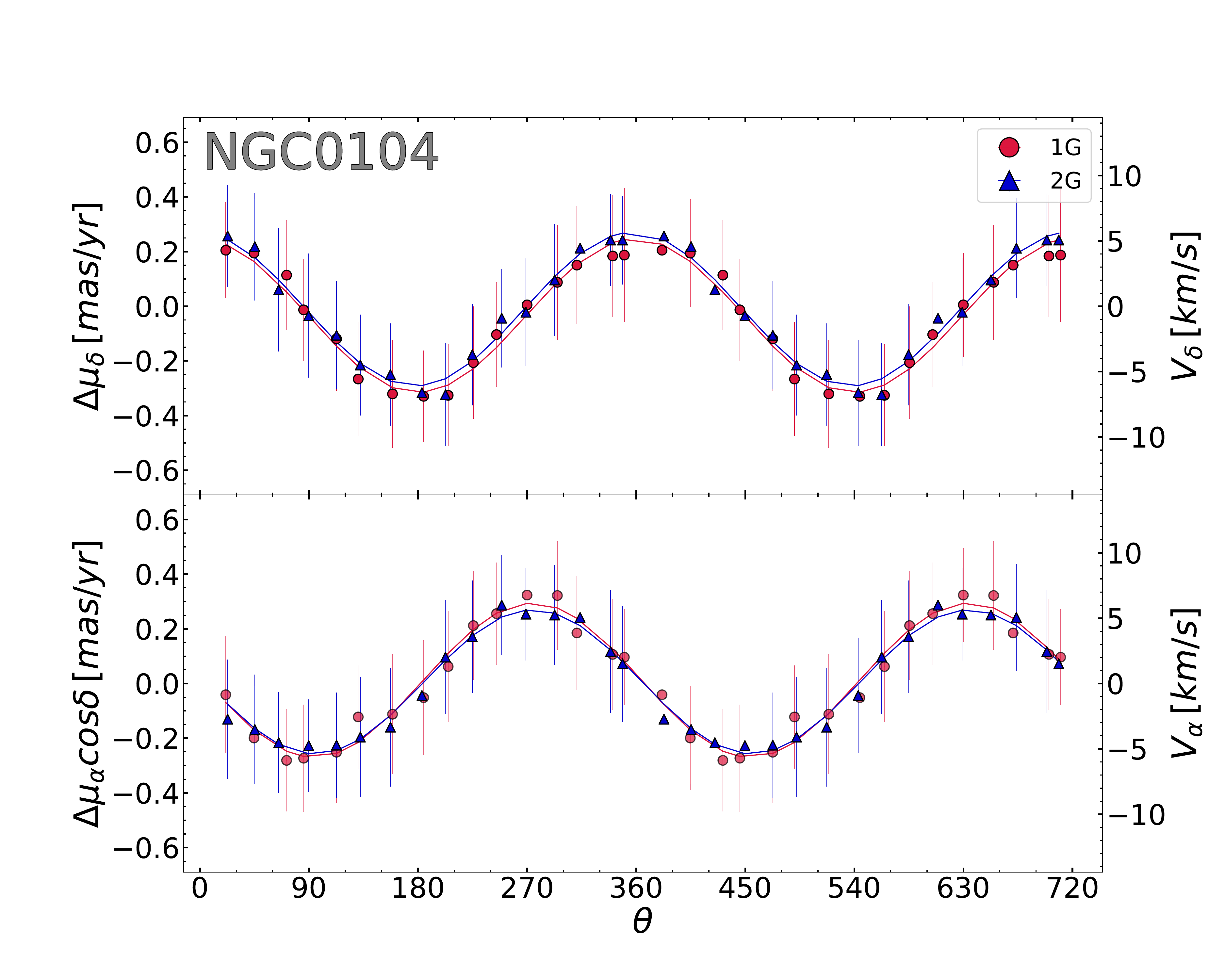} 
  \includegraphics[width=6.5cm]{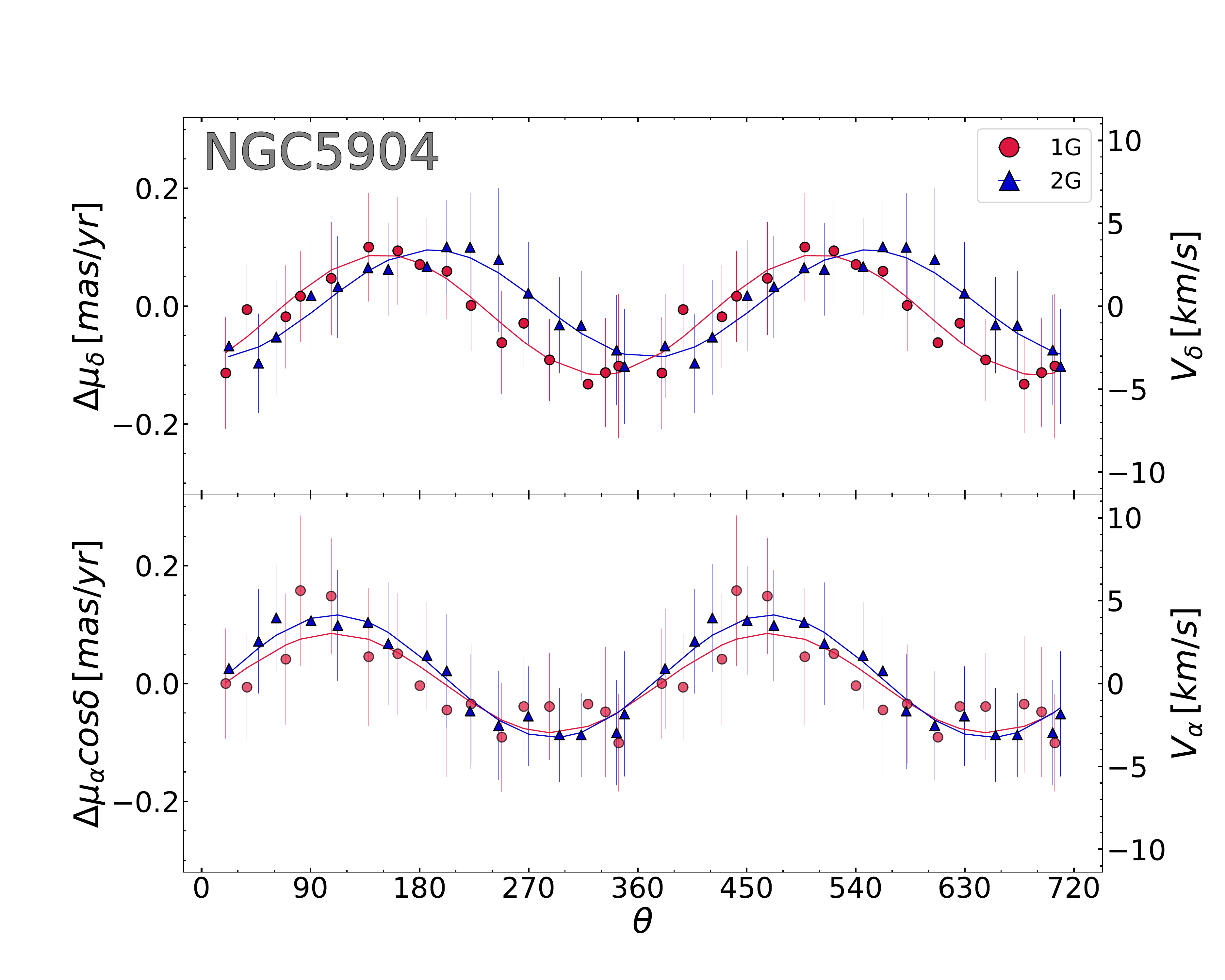} 
  \caption{Rotation curve for the 1G and 2G stars in NGC\,0104 (left panel) and NGC\,5904 (right panel).}
  \label{fig: rot}
\end{figure}

\section{Method and results}
\subsection{Rotation}
  To study the kinematics of multiple stellar populations over a large field of view, we exploit the pseudo color $C_{\rm U,B,I}=(U-B)-(B-I)$, which provides an efficient tool to identify stellar populations with different light-element abundance along the RGB,  to identify 1G and 2G stars.  \\
  As a first step, we determined cluster memberships of each star exploiting Gaia DR2 accurate proper motions, parallaxes and astrometric quality parameter, and we corrected our photometry for differential reddening. \\
  For each cluster, we studied the internal rotation of 1G and 2G stars on the plane of the sky by means of Gaia DR2 proper motions. 
  Figure~\ref{fig: rot} shows the median motions in different angular regions for 1G and 2G stars in NGC 0104 and NGC 5904, which are the two clusters with clear evidence of rotation.  Interestingly, the two populations of NGC\,5904 exhibit different rotation, and in particular the phase shift between the rotation curves of 1G and 2G stars in the $\mu_{\rm \delta}$ component is significant at 3$\sigma$ level. On the other hand, NGC\,0104 is consistent with two populations sharing the same rotation. However, some hints of different rotations between 1G and 2G stars are observed in the external regions of this cluster.
  There is no evidence for rotation in 1G and 2G stars of the remaining clusters (NGC\,0288, NGC\,6121, NCG\,6254, NGC\,6752 and NGC\,6838).

\subsection{Internal kinematics}
\begin{figure}[h!]
  \centering
  \includegraphics[width=6cm]{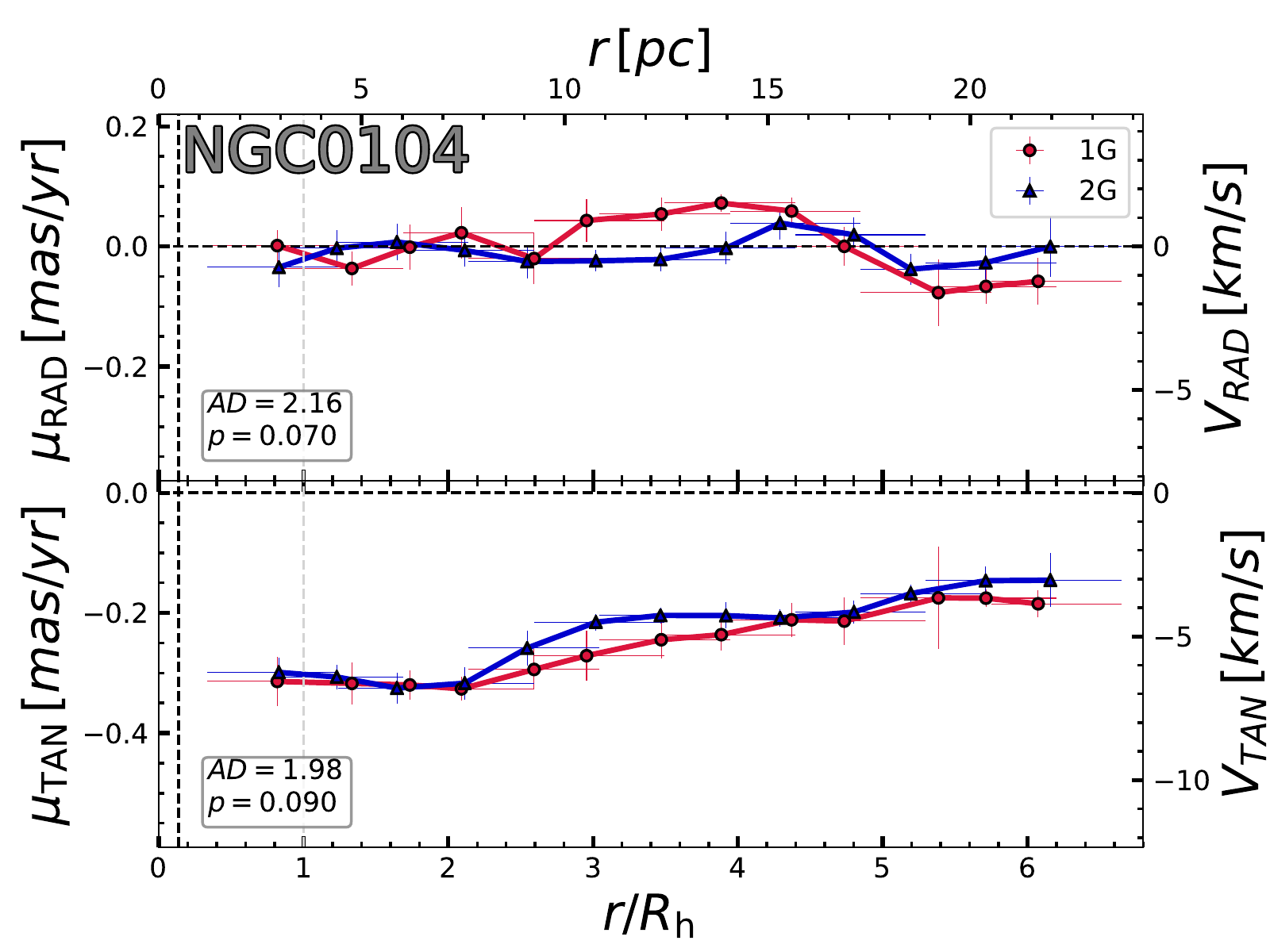} 
  \includegraphics[width=6cm]{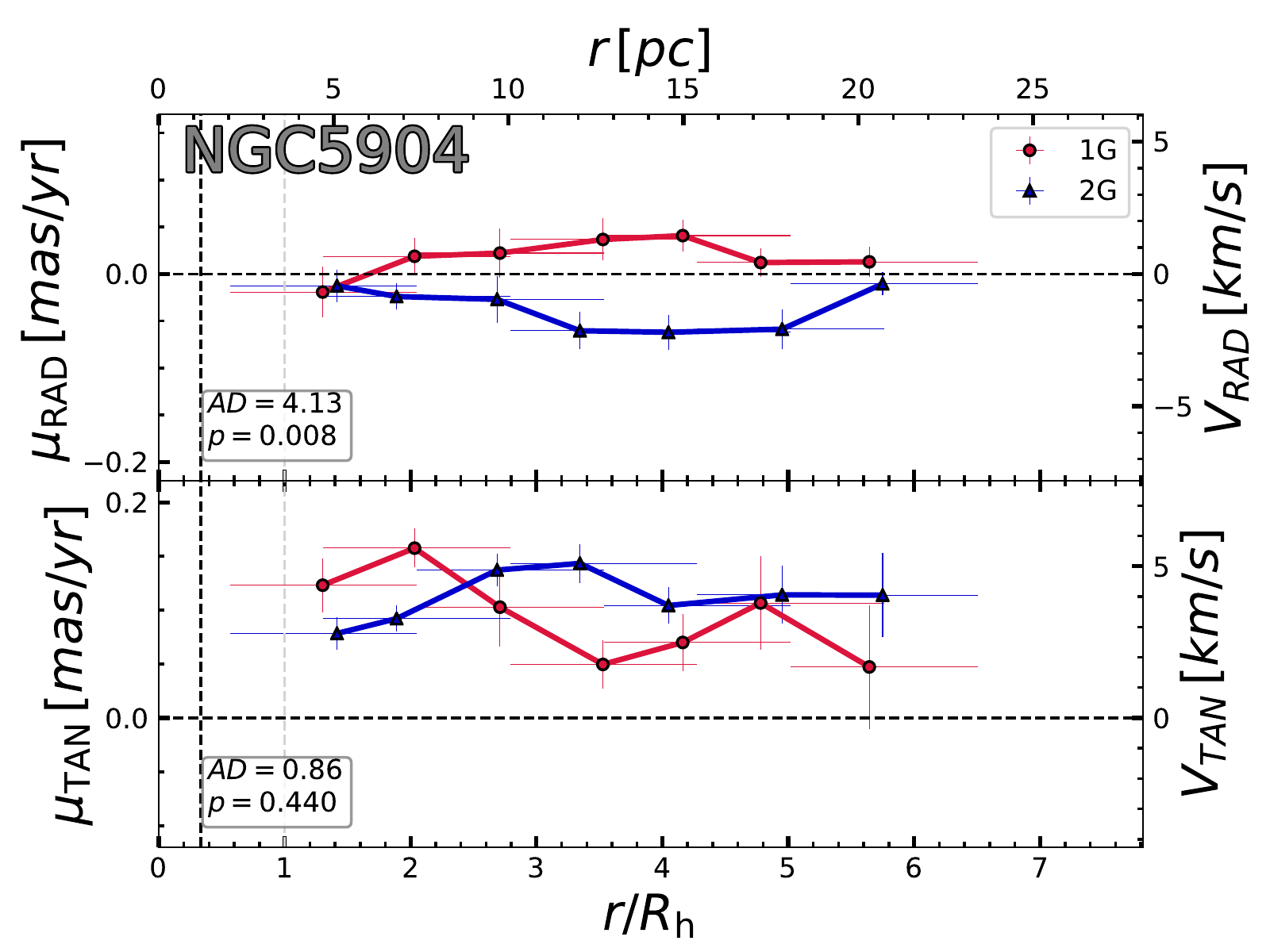} 
  \includegraphics[width=6cm]{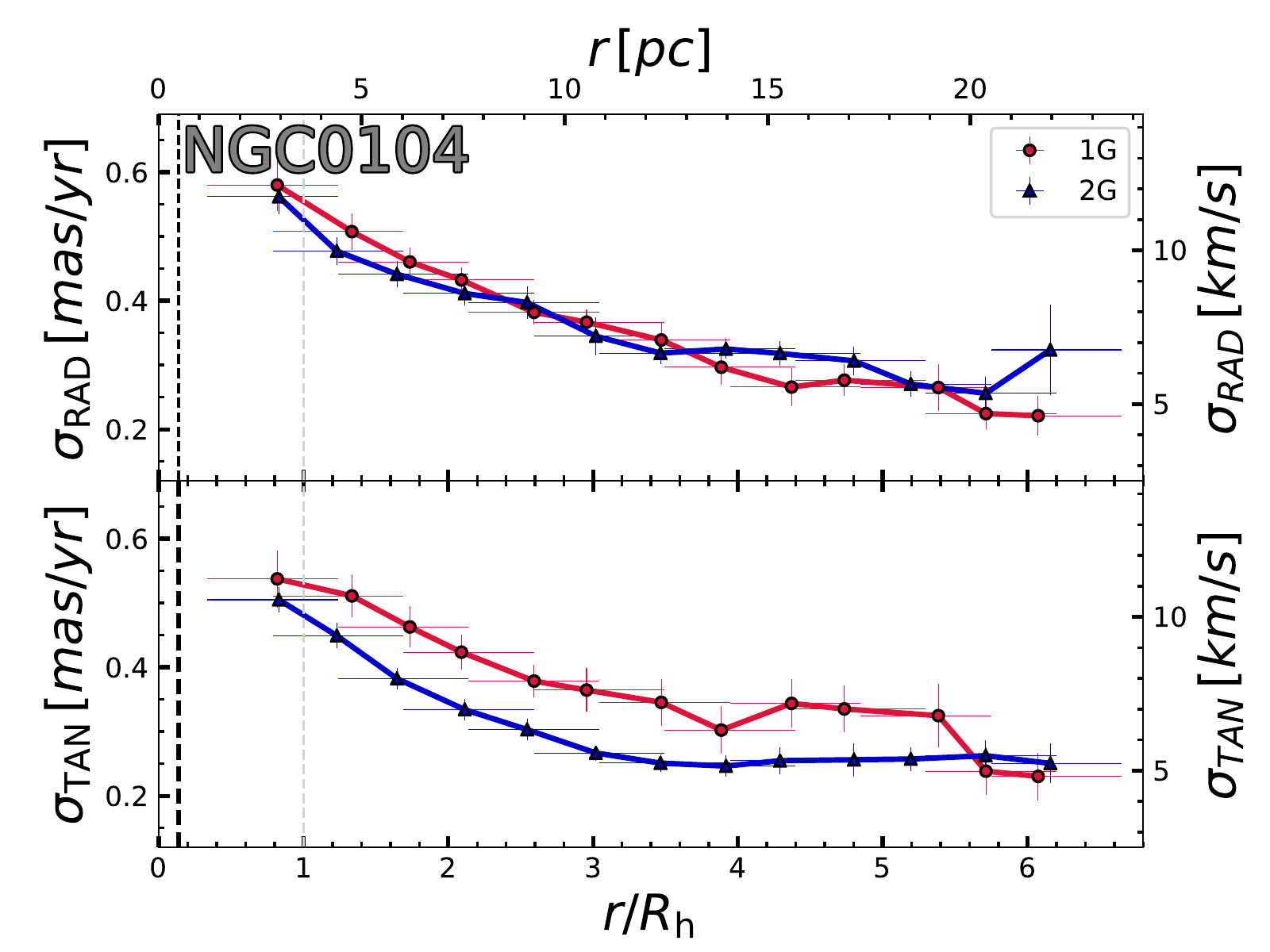} 
  \includegraphics[width=6cm]{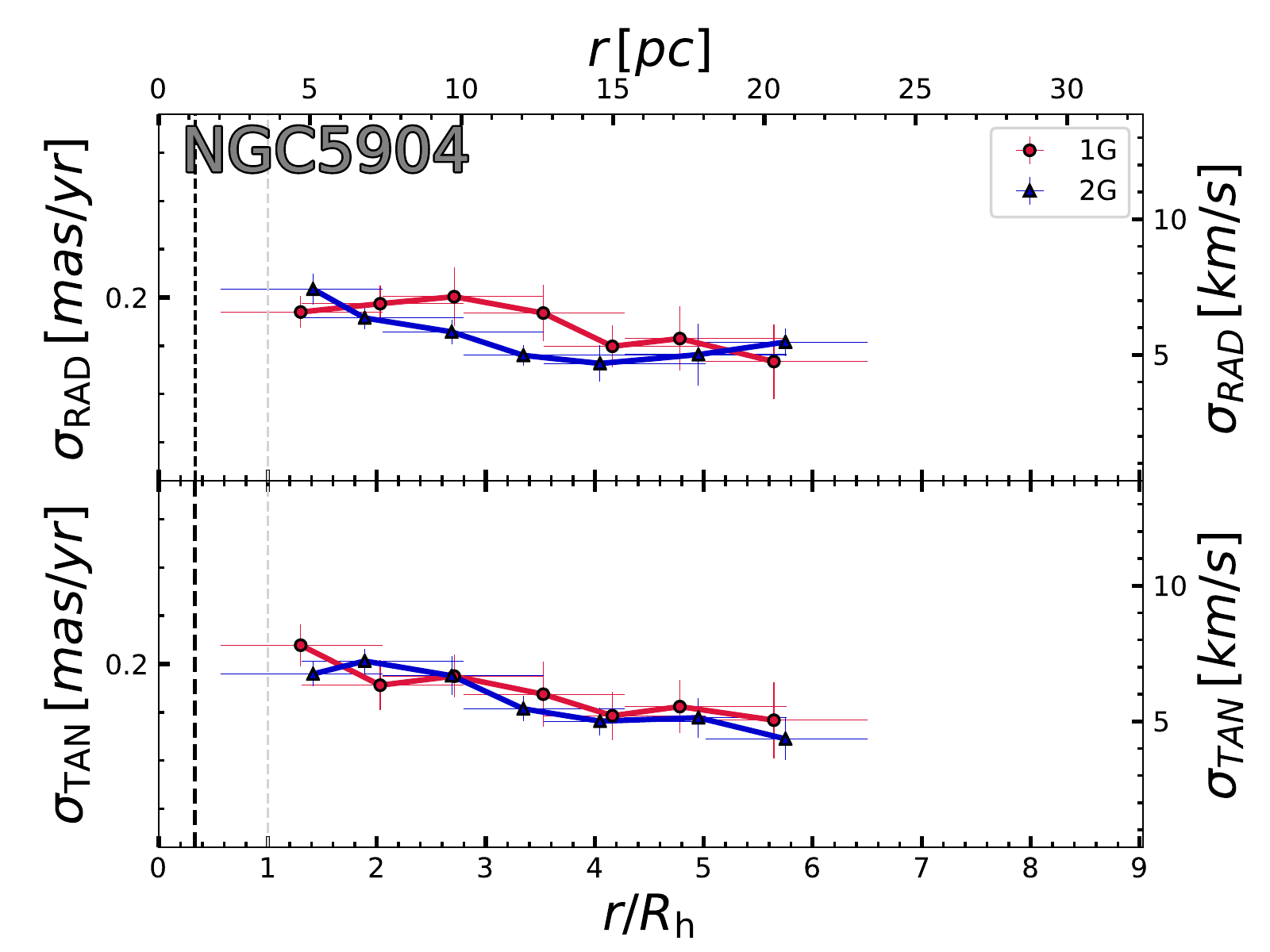} 
  \includegraphics[width=6cm, height=4.8cm]{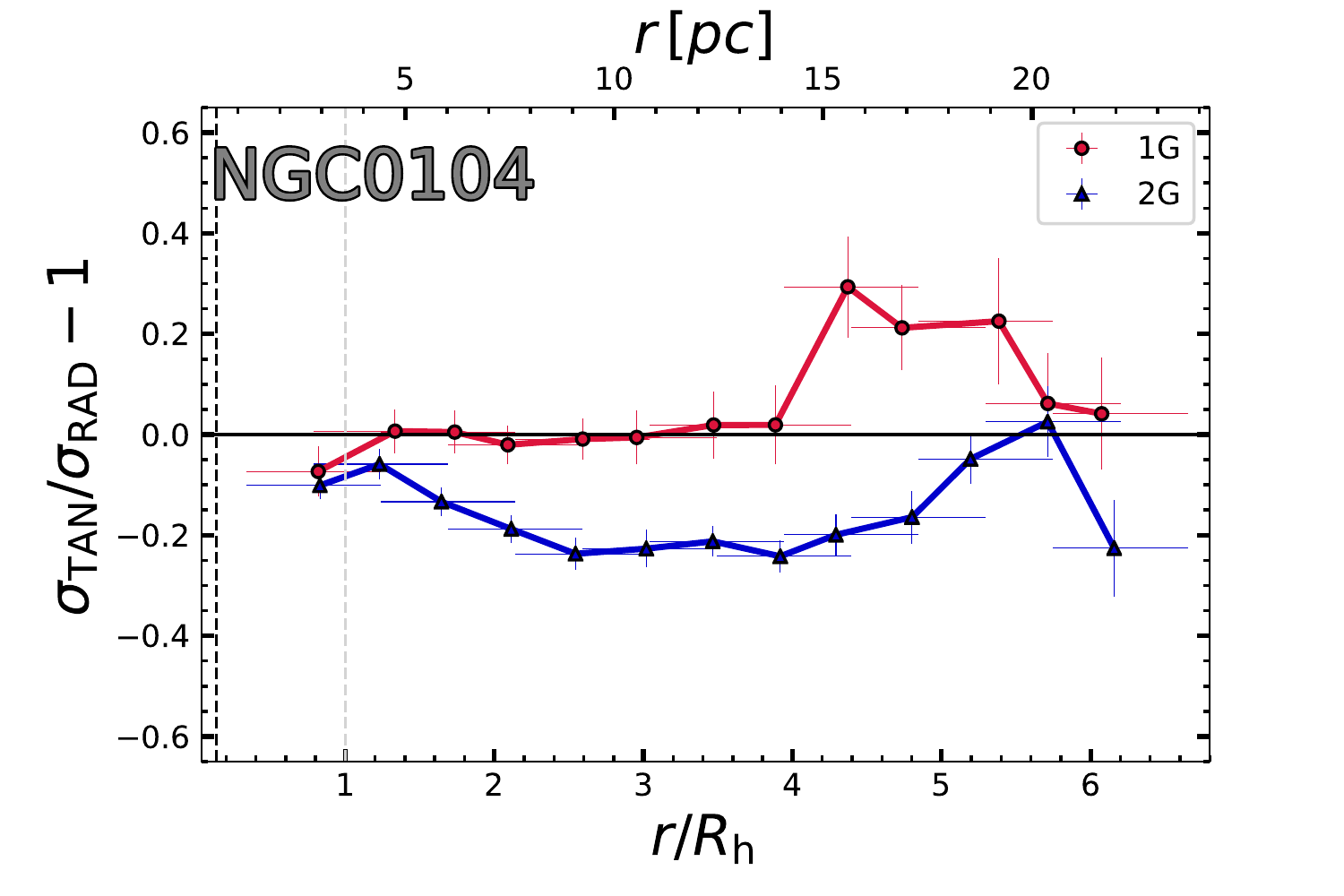} 
  \includegraphics[width=6cm, height=4.8cm]{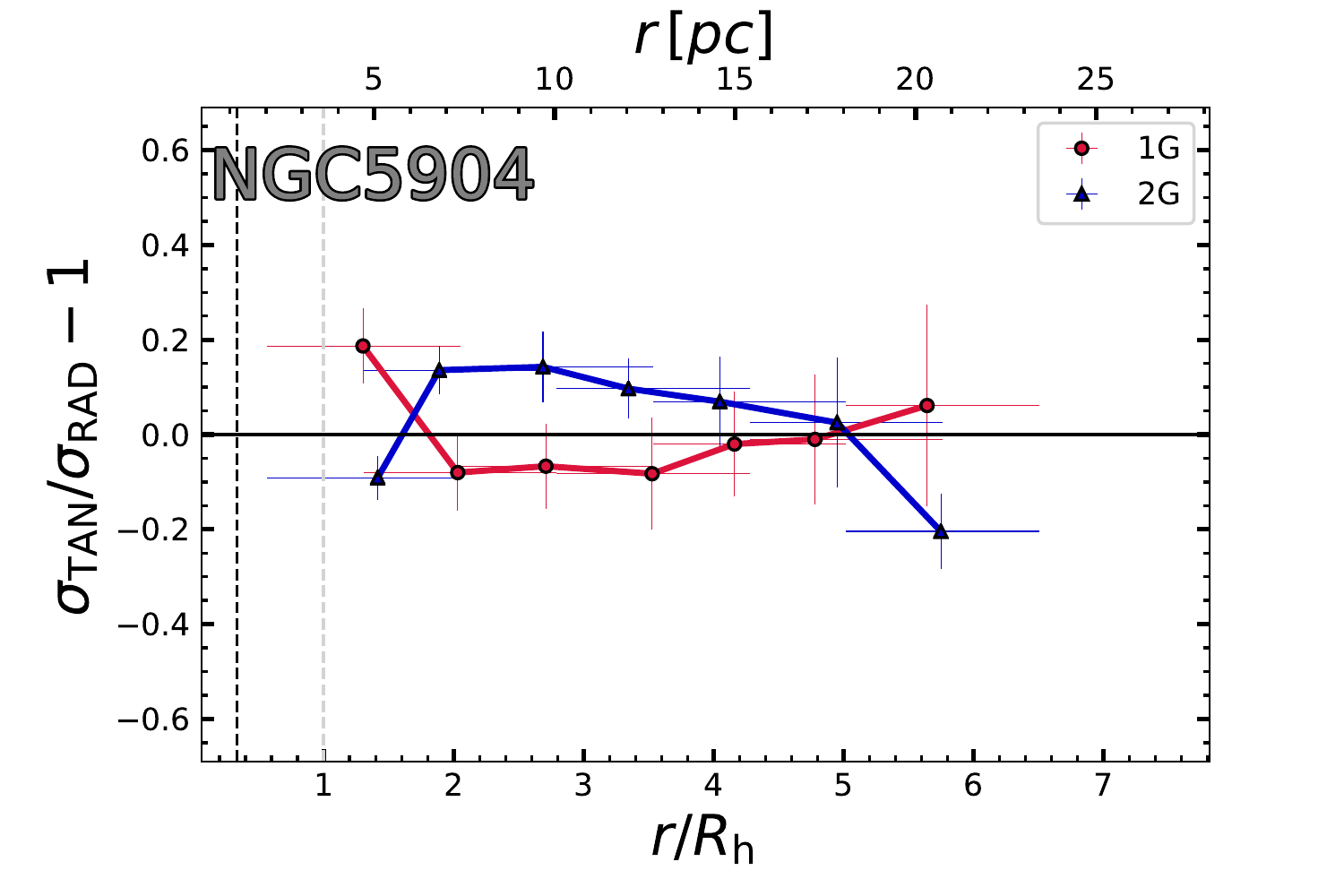} 
  \caption{Median profile, velocity dispersion profile and anisotropy profile of NGC\,0104 (left panels) and NGC\,5904 (right panels).}
  \label{fig: profile}
\end{figure}

  To study the internal motion of stars as a function of the radial distance from the cluster center we divided the cluster field into different circular annuli, and for each one we computed the median radial ($\mu_{\rm RAD}$) and tangential ($\mu_{\rm TAN}$) components of proper motions for 1G and 2G stars, on the plane of the sky.
  We then derived the velocity dispersion profiles, $\sigma_{\rm RAD},\,\sigma_{\rm TAN}$, and the anisotropy profile $\sigma_{\rm TAN}/\sigma_{\rm RAD}-1$, shown in Figure~\ref{fig: profile}.\\
  Most of the GCs show that their two main populations share similar velocity profiles and any difference between the velocities of 1G and 2G stars is smaller than $\sim$1 km/s. 
  NGC\,5904 is a remarkable exception. Indeed, between $\sim$2 to $\sim$5 half-light radii $(R_{\rm h})$ from the center, 1G stars exhibit higher radial motions than 2G stars. NGC\,0104 exhibits a similar, but milder, trend. Small tangential velocity differences $(\lesssim 1\,{\rm km/s})$ between 2G and 1G stars are also present in NGC\,6254 and NGC\,6752. \\
  We confirm that NGC\,0104 exhibits strong differences in the degree of anisotropy of the two populations, with the 2G being more radially anisotropic than the 1G.  
  Besides NGC\,0104, also NGC\,6752 and NGC\,6838 show hints of a more radially anisotropic 2G within $\sim$1-2 $R_{\rm h}$ from the cluster center. 
  On the other hand, our results suggest a more tangential anisotropic 1G within $\sim$1$R_{\rm h}$ for NGC\,6121.  The remaining clusters are consistent with being isotropic stellar systems.

  All our findings constitute strong constraints for existing and future multiple population scenarios. Self-enrichment scenarios, and in particular the AGB scenario, seem to be able to produce different spatial distribution and kinematics between the first and second generation. This scenario, which is the one that have been studied more in detail in terms of dynamics, predicts a higher central concentration for 2G with respect to 1G stars. 1G stars have higher velocity dispersion compared to 2G stars and they show a smaller amount of radial anisotropy. However, all these signatures could be washed out by the two-body relaxation of the clusters (\cite[Mastrobuono-Battisti \& Perets 2016]{mastrobuono2016}). \\
  This work constitutes the first coherent analysis of the dynamics of multiple stellar populations over a wide field of view, and adds new important constraints to our understanding of the multi-populations phenomenon.
\section*{Acknowledgments} 
\small
  This work has received funding from the European Research Council (ERC) 
  under the European Union's Horizon 2020 research innovation programme 
  (Grant Agreement ERC-StG 2016, No 716082 'GALFOR', PI: Milone, http://progetti.dfa.unipd.it/GALFOR), 
  and the European Union's Horizon 2020 research and innovation programme 
  under the Marie Sk\l odowska-Curie (Grant Agreement No 797100, beneficiary: 
  Marino). APM and MT acknowledge support from MIUR through the FARE project 
  R164RM93XW SEMPLICE (PI: Milone).  AMB acknowledges support by Sonderforschungsbereich (SFB) 881 ‘The Milky Way System’ of the German Research Foundation (DFG).

\end{document}